%% file: nikolaev_chervon_en.tex
\documentclass[fleqn,11pt]{article}

\usepackage{amsmath,amssymb} 

\input preamble1.tex

\begin{document}
\twocolumn[
\jnumber{2}{2016}

\Title {The effect of  universe inhomogeneities on cosmological distance\yy measurements}

\Aunames{A. V. Nikolaev\auth{a,1} and S. V. Chervon\auth{a,b}}

\Addresses{
\addr a {Ilya Ulyanov State Pedagogical University, 432700 Ulyanovsk, Russia}
\addr b {Astrophysics and Cosmology Research Unit, School of Mathematics,\\ 
	Statistics and Computer Science, University of KwaZulu-Natal, Private Bag X54 001, 
	Durban 4000, South Africa}
	}

\Rec{January 21, 2016}

\Abstract
{Using the focusing equation, the equation for the cosmological angular diameter distance\foom 2 	
is derived, based on the ideas of Academician Zel'dovich, namely, that the distribution of matter at small
angles is not homogeneous, and the light cone is close to being empty. We propose some ways of testing
a method for measuring the angular diameter distances and show that the proposed method leads to
results that agree better with the experimental data than those obtained by the usual methods.}

\bigskip\bigskip

] 
\email 1 {ilc@xhns.org}
\foox 2 {Not to be confused with the ``angular distance'' defined as a distance between 
	 points on the celestial sphere and measured in radians or degrees.}

 The abundance of observational data in modern cosmology allows for testing a number of ideas put 
 forward in the past. One of such ideas is the approach of Academician Ya. B. Zel'dovich to 
 measurements of the cosmological angular diameter distances [1], which takes into account 
 that null geodesics propagate in a homogeneous Friedmann universe, but the null geodesic 
 congruence (or light cone) from the source experiences a smaller focusing than in a homogeneous
 universe.\footnote
	{A light cone is understood here and henceforth as ``a cone of light rays'' or a cone 
	  that bounds a beam of null geodesics, not to be confused with the light (or null) cone of
	  relativity theory, which separates spacelike and timelike directions.} 
 Such an effect is possible if the density of matter inside a light cone 
 is smaller than the mean density in the Friedmann universe .

 Ref. [2] suggested a derivation of a generalized differential equation using such tools as the null
 geodesics  and the ratio of longitudinal and transverse angular momentum of a photon. We will 
 show that this equation  can also be obtained on the basis of the focusing equation [6], which 
 follows from the Sachs equations [3] (a special case of the Raychaudhuri equations [5]):
\beq
	\frac{d^2}{d\lambda^2}\sqrt{S} =
	 -\left(|\sigma|^2 + \frac{1}{2}R_{\alpha\beta}k^\alpha k^\beta\right)\sqrt{S},
\label{nik-focusingeq}
\eeq
 where $S$ is the light cone cross-section area, $\lambda$ is the affine parameter, $k^\alpha$ is the null
 wave vector, Greek indices run over the values $(0,1,2,3)$, and $\sigma$ is the shear defined as follows:
\beq
	|\sigma|^2 = \frac{1}{2}k_{\alpha;\beta}k^{\alpha;\beta}-\frac{1}{4}(k^\alpha_{;\alpha})^2.
\label{nik-shear}
\eeq

 In the Friedmann-Robertson-Walker metric 
\beq
	ds^2 = dt^2 - a(t)^2[dr^2 + f^2(r)(d\theta^2 + \sin^2\theta d\phi^2)]
\label{nik-metric}
\eeq
 the wave vector has the following components for the arriving geodesics: $k^\alpha_{in} = 
 (-1/a, 1/a^2,0,0)$; the affine parameter is related to time through the scale factor:
 $d \lambda = - a\,dt$. Directly calculating the covariant derivatives in the metric \eqref{nik-metric},  
 we verify that $|\sigma|^2=0$: 
\beq
      |\sigma|^2 = \frac{\left(f\dot{a} - f'\right)^2}{f^2a^4} -  \frac{\left(f\dot{a} - f'\right)^2}{f^2a^4} = 0.
\label{nik-sigma}
\eeq
 Expressing the light cone cross-section in terms of the linear size of the source $S = \frac{\pi l^2}{4}$
 (see [2] for details) and substituting \eqref{nik-sigma} into \eqref{nik-focusingeq}, we obtain
\beq
	\ddot{l} - \frac{\dot{a}}{a}\dot{l} + a^2\frac{1}{2}R_{\alpha\beta}k^\alpha k^\beta l = 0.
\label{nik-zeq}
\eeq
 Using in  \eqref{nik-zeq} the definition of an angular diameter distance $d_a = l/\phi$, we arrive at
\beq
      \ddot{d_a} - \frac{\dot{a}}{a}\dot{d_a} + a^2\frac{1}{2}R_{\alpha\beta}k^\alpha k^\beta d_a = 0.
\label{nik-daeq}
\eeq

 Contracting the Einstein equations
\beq
	R_{\alpha\beta}-\frac{1}{2}Rg_{\alpha\beta}=\kappa T_{\alpha\beta}
\label{nik-einstein}
\eeq
 with $k^\alpha k^\beta$ (using the null nature of the wave vector, $g_{\alpha\beta}k^\alpha k^\beta = 
 0$), we obtain
\beq
	\frac{1}{2}R_{\alpha\beta}k^\alpha k^\beta = \frac{\kappa}{2}T_{\alpha\beta}k^\alpha k^\beta.
\label{nik-Ralphabeta2}
\eeq
 Here $\kappa$ is the Einstein gravitational constant.
 It should be noted that the beam focusing is affected by only the local value of the Ricci tensor, or, 
 due to the Einstein equations, by the local value of the energy-momentum tensor. 
 The result \eqref{nik-Ralphabeta2}  allows us to convert the equation for the cosmological 
 angular diameter distance \eqref{nik-daeq} to the form
\beq
	\ddot{d_a}-\frac{\dot{a}}{a}\dot{d_a}+4\pi Ga^2T_{\alpha\beta}k^\alpha k^\beta d_a = 0,
\label{nik-daeq2}
\eeq
 where $T_{\alpha\beta}$ is the local value of the energy-momentum tensor inside the light cone. 
 One can introduce the parameter $\alpha$ showing how much matter is there inside the cone:
\beq
	\alpha = \frac{T_{\alpha\beta}k^\alpha k^\beta}{T_{\alpha\beta}^{\rm full}k^\alpha k^\beta}.
	\label{nik-alpha}
\eeq
 Using the definition of the energy-momentum tensor for a perfect fluid, $T_\alpha^\beta = 
 \diag (\rho,-p,-p,-p)$, we obtain:
\beq
	a^2T_{\alpha\beta}k^\alpha k^\beta = p+\rho.
\label{nik-focpf}
\eeq
 Making explicit the components for the  $\Lambda$CDM model, we get:
\beq
	p+\rho = p_\Lambda + p_M + p_R +\rho_\Lambda + \rho_M + \rho_R.
\label{nik-pplusrho}
\eeq
 Using the equations of state for baryonic matter ($p_M=0$), dark energy 
 ($p_\Lambda=-\rho_\Lambda$), and radiation ($p_R=\rho_R/3$), we convert the relation 
\eqref{nik-pplusrho} to
\beq
	p+\rho = \frac{4}{3}\rho_R +\rho_M,
\label{nik-pplusrho2}
\eeq
 where
\bearr
	\rho_M = \frac{3H_0^2\Omega_M}{8\pi G}\left(\frac{a_0}{a}\right)^3,  
\nnn
	\rho_R = \frac{3H_0^2\Omega_R}{8\pi G}\left(\frac{a_0}{a}\right)^4.  \label{nik-rhoM}
\ear
 Using \eqref{nik-focpf} and \eqref{nik-pplusrho2}, \eq \eqref{nik-daeq2} acquires the form
\beq
	\ddot{d_a}-\frac{\dot{a}}{a}\dot{d_a}+4\pi G\left(\frac{4}{3}\rho_R+\rho_M\right) d_a = 0
\label{nik-daeq3}
\eeq
 Thus it has been established that dark energy does not participate in focusing of the light rays 
 (which makes clear the question raised in [11]). Since inside the light cone, as a rule, $\rho_M$ and
 $\rho_R$ tend to zero, the value of $d_a$ will be larger than in Friedmann's homogeneous  model [2].

 Let us now discuss a number of tests for approaches to calculations of the angular diameter distance
 which follow from the data on the Sunyaev-Zel'dovich effect (SZE) for galaxy clusters [12--14].
 The angular diameter distance may be expressed through the SZE data [12]:  
\bearr
    d_a^{\rm SZE} = \frac{(\Delta T_0)^2}{S_{X0}}\left(\frac{m_e c^2}{k_B T_{e0}}\right)^2 
\nnn\times
    \frac{\lambda_{eH0}\mu_e/\mu_H}{4\pi^{3/2}f^2_{(x,T_e)}T^2_{\rm CMB}
	\sigma^2_T(1+z)^4}\frac{1}{\theta_c}x
\nnn\times
    \left[\frac{\Gamma(3\beta/2)}{\Gamma(3\beta/2-1/2)}\right]^2 
	\frac{\Gamma(3\beta-1/2)}{\Gamma(3\beta)},
\ear
  where $\Gamma(x)$ is the gamma function, $S_{X0}$ is the X-ray surface brightness of the 
  cluster center, $z$ is the redshift, $\lambda_{eH}$ is the cooling function of the cluster center, 
  $\sigma_T$ is the total scattering cross-section, $k_B$ is the Boltzmann constant, $\Delta T_0$ 
  is the SZE temperature difference, $\theta_c$ is the angular size of the galactic nucleus, $m_e$ is 
  the electron mass, $f(x,T_e)$ is the SZE frequency dependence, and $T_{\rm CMB}$ 
  is the temperature of the microwave background radiation.

  Thus there emerges a test for the angular diameter distance connected with the Hubble 
  constant $H_0$. Let us write down the solution of \eqref{nik-daeq3} for an empty light cone in 
  $\Lambda$CDM [2]:
\beq
    d_a^{\rm empty} = \frac{1}{H_0}\int_\frac{1}{1+z}^1 \frac{dx}{\sqrt{\Omega_S}}
\label{nik-emptysol}
\eeq
  where $\Omega_S = \Omega_\Lambda + \Omega_kx^{-2}+\Omega_Mx^{-3}+\Omega_Rx^{-4}$,
  while for a light cone filled with matter whose density is equal to the mean density of the Universe,
  the solution of \eqref{nik-daeq3} has the form 
\bearr
	d_a^{\rm full} = 
	\frac{1}{1+z}\frac{1}{H_0\sqrt{\Omega_k}} \sin\int_\frac{1}{1+z}^1\sqrt{\Omega_k}
	\frac{dx}{x^2\sqrt{\Omega_S}}
\nnn \inch
	{\rm for}\ \  k=1,
\nnn
	d_a^{\rm full} =
	\frac{1}{1+z} \int_\frac{1}{1+z}^1\frac{dx}{H_0x^2\sqrt{\Omega_S}} \quad {\rm for} \ \ k=0,
\nnn
	d_a^{\rm full} =	
	\frac{1}{1+z} \frac{1}{H_0\sqrt{\Omega_k}}\sinh\int_\frac{1}{1+z}^1\sqrt{\Omega_k}
	\frac{dx}{x^2\sqrt{\Omega_S}}, 
\nnn \inch
	{\rm for}\ \  k= -1. 			\label{nik-fullsol}
\ear
  This allows us to compare the values of the Hubble constant predicted by the standard solution  
  \eqref{nik-fullsol} and the new formula \eqref{nik-emptysol}. The value of the cosmological 
  angular diameter distance is calculated directly from the SZE, which means that equating 
  $d^{\rm empty}_a$ and $d_a^{\rm full}$ to $d_a^{\rm SZE}$, we can find $H_0$. Therefore, 
  for an empty light cone we obtain
\beq
     H_0^{\rm empty} = \bigg(\int_\frac{1}{1+z}^1 \frac{dx}{\sqrt{\Omega_S}}\bigg)
	\bigg/d_a^{\rm ZSE},				\label{nik-H0empty}
\eeq
  while for a full light cone
\bearr
    H_0^{\rm ZSE} = 
	 \frac{1}{(1+z)d_a^{ZSE}}\frac{1}{\sqrt{\Omega_k}} \sin\int_\frac{1}{1+z}^1
	\sqrt{\Omega_k}\frac{dx}{x^2\sqrt{\Omega_S}}
\nnn \inch
	{\rm for}\ \  k=1,
\nnn
    H_0^{\rm ZSE} = 
	\frac{1}{(1+z)d_a^{ZSE}} \int_\frac{1}{1+z}^1\frac{dx}{x^2\sqrt{\Omega_S}} 
		 \quad {\rm for} \ \ k=0,
\nnn
    H_0^{\rm ZSE} = 		    
	\frac{1}{(1+z)d_a^{ZSE}} \frac{1}{\sqrt{\Omega_k}}\sinh\int_\frac{1}{1+z}^1
	\nhq \sqrt{\Omega_k}\frac{dx}{x^2\sqrt{\Omega_S}}
\nnn \inch
	{\rm for}\ \  k=-1.			  \label{nik-H0full}
\ear

 The calculation of the simple averages from the data for clusters of galaxies [12] allows us to 
 conclude that more consistent values of the Hubble constant are given by \eqs \eqref{nik-H0empty} 
 than \eqref{nik-H0full}. A detailed analysis of the galactic cluster data in the context of using 
 \eqs \eqref{nik-H0empty} can serve as a material for further experimental studies.

  The next test is connected with the duality between the cosmological angular diameter 
  distance $d_a$ and the luminosity distance $d_l$: 
\beq
	\eta = \frac{d_l}{d_a}(1+z)^{-2} = 1,	\label{nik-edington}
\eeq
  which follows from the Eddington identity [4]:
\beq
          r_s^2 = r_o^2(1+z)^2,
\label{nik-rrel}
\eeq
  where  $r_s$ is the distance to the source and $r_o$ is the distance to the observer,
  which is determined through the solid angle and the cross-section area, $dS=r^2d\Omega$ 
  (for more details see [7]).

  In [10], an attempt is undertaken to test the validity of the identity \eqref{nik-edington}
  on the basis of the data from galaxy clusters [12] using the formula
\beq
	\eta(z) = \sqrt{\frac{d_a^{\rm Th}}{d_a^{\rm data}}},
\label{nik-adf5}
\eeq
  where $d_a^{\rm Th}$ is obtained from theoretical calculations according to \eqref{nik-emptysol} 
  or \eqref{nik-fullsol}. An analysis [9] shows that the new method of calculations of the angular 
  diameter distance allows one to experimentally confirm the identity \eqref{nik-edington}
  with a greater accuracy than the standard method. 

  Another approach to verification of the identity has been proposed in [8], using the  
  surface brightness data in the X-ray spectrum together with SZE data [13, 14]. To assess 
  the validity of the identity \eqref{nik-edington}, one uses the mass fraction of gas in the galaxy, 
  $f=M_{\rm gas}/M_{\rm Tot}$, and the ratio
\beq
	\eta(z) = \frac{f_{\rm SZE}}{f_{\rm X-ray}},
\label{nik-etafunc}
\eeq
  where $f_{\rm SZE}$ is the mass fraction of gas measured with the aid of the SZE,  and 
 $f_{\rm X-ray}$  is the same calculated  assuming the validity of  \eqref{nik-edington} [14, 15].  

  If we insert the correction connected with applying the new method of calculation of the angular 
  diameter distance \eqref{nik-emptysol}, an analysis shows that the $z$ dependence of $\eta(z)$'
  becomes closer to unity. This argues in favor of the new method of calculation of the angular
  diameter distance. The very distribution of values of $\eta$ is shifted to unity, showing that the duality
  identity for cosmological distances holds with an accuracy of $1\sigma$, in contrast to the result
 ($2\sigma$) of the original work [8]..

  In conclusion, we would like to note that, based on the aforementioned reasoning, Zel'dovich's idea
  receives a confirmation. In contrast to the papers [16, 17], developing the ideas of Dyer and Roeder,
  we obtain simpler calculation formulas which reflect the physical meaning of measurements of the  
  cosmological angular diameter distance in the Friedmann universe taking into account the
  inhomogeneities. Our approach makes it possible to pass on to the stage of experimental verification.

\subsection*{Acknowledgments}

 The authors are grateful to the participants of the Zel'manov seminar on gravitation and cosmology at
 the Sternberg astronomical Institute of Moscow State University and the VNIIMS gravitational seminar 
 for their constructive criticism and helpful discussions.
 SC and AN were partly supported by the State order of Ministry of education and science of RF number  2014/391 on the project 1670.

\end{document}

%% file: preamble1.tex
\usepackage{amsfonts,amssymb,cite}
\usepackage{graphicx}

\def\noi{\noindent}
\def\jnumber#1#2{\thispagestyle{empty} \noi\unitlength=1mm
    	\begin{picture}(178,10)
            \put(177,15){\llap{\large\it Grav. Cosmol. No.\,#1, #2}}
                    \end{picture}}

\newcommand{\Title}[1]{\noi {{\Large\bf #1}}\\[1ex]}

\def\Aunames#1{\noi{\bf #1}}
\def\auth#1{${}^{#1}$}
\def\Addresses#1{\medskip\noi \protect
	\begin{description}\itemsep -3pt {\it #1} \end{description}}
\def\addr#1#2{\item[${}^{#1}$]{\it #2}}

\newcommand{\Rec}[1]{\noi {\it Received #1} \\}

\newcommand{\Abstract}[1]{\vskip 2mm \begin{center}
        \parbox{16.4cm}{\small\noi #1} \end{center}\medskip}

\newcommand{\foom}[1]{\protect\footnotemark[#1]}
\newcommand{\foox}[2]{\footnotetext[#1]{#2}\addtocounter{footnote}{1}}
\def\email#1#2{\footnotetext[#1]{e-mail: #2}\addtocounter{footnote}{1}}

\def\nqq{\hspace*{-2em}}
\def\nhq{\hspace*{-0.5em}}

\def\inch{\hspace*{1in}}

\def\Jl#1#2{#1 {\bf #2},\ }

\def\ApJ#1 {\Jl{Astroph. J.}{#1}}
\def\CQG#1 {\Jl{Class. Quantum Grav.}{#1}}
\def\DAN#1 {\Jl{Dokl. AN SSSR}{#1}}
\def\GC#1 {\Jl{Grav. Cosmol.}{#1}}
\def\GRG#1 {\Jl{Gen. Rel. Grav.}{#1}}
\def\JETF#1 {\Jl{Zh. Eksp. Teor. Fiz.}{#1}}
\def\JETP#1 {\Jl{Sov. Phys. JETP}{#1}}
\def\JHEP#1 {\Jl{JHEP}{#1}}
\def\JMP#1 {\Jl{J. Math. Phys.}{#1}}
\def\NPB#1 {\Jl{Nucl. Phys. B}{#1}}
\def\NP#1 {\Jl{Nucl. Phys.}{#1}}
\def\PLA#1 {\Jl{Phys. Lett. A}{#1}}
\def\PLB#1 {\Jl{Phys. Lett. B}{#1}}
\def\PRD#1 {\Jl{Phys. Rev. D}{#1}}
\def\PRL#1 {\Jl{Phys. Rev. Lett.}{#1}}

\def\lal{&&\nqq {}}
\def\eq{Eq.\,}
\def\eqs{Eqs.\,}
\def\beq{\begin{equation}}
\def\eeq{\end{equation}}
\def\bear{\begin{eqnarray}}
\def\bearr{\begin{eqnarray} \lal}
\def\ear{\end{eqnarray}}
\def\earn{\nonumber \end{eqnarray}}

\def\nnn{\nonumber\\ \lal }

\def\yy{\\[5pt] {}}

\def\diag{\mathop{\rm diag}\nolimits}

